\documentstyle{mn}

\input psfig

\title[Spiral structure in IP Pegasi]{Simulations of spiral 
	structure in the accretion disc \\
	of IP Pegasi during outburst}
	
\author[P.J. Armitage \& J.R. Murray]{P.J. Armitage$^1$ and J.R. Murray$^2$ \\
 	$^1$ Canadian Institute for Theoretical Astrophysics, McLennan Labs,
	60 St George St, Toronto, M5S 3H8, Canada \\
	$^2$ The Astrophysical Theory Centre, 
	Australian National University, ACT 0200, Australia}	
 	
\begin{document}

\maketitle

\begin{abstract} 	
We consider the implications of the detection of spiral structure
in the accretion disc of the binary IP Pegasi. We use numerical simulations
of the development of a disc outburst to construct predicted Doppler 
tomograms, which are found to be in close agreement with the observations
if the spiral pattern arises as a transient feature when the disc expands
viscously at the start of the outburst. The good agreement
of such viscous disc simulations with the data is consistent with models 
in which most of the angular momentum transport in the disc originates in 
internal stresses rather than globally excited waves or shocks. Future
detailed observations of the development of transient spiral features offer 
the potential to measure the dependence of the disc viscosity on the local 
physical conditions in the disc. 
\end{abstract}


\section{INTRODUCTION}
Recent observations of the dwarf nova IP Pegasi provide 
convincing evidence for spiral structure in the emission 
from an accretion disc in a binary system (Steeghs, Harlaftis \& Horne 1997). 
During an outburst, changes in the profile of spectral lines with 
binary phase were inverted using the technique of Doppler tomography
(Marsh \& Horne 1988) to reveal a loosely wrapped, two-armed spiral
pattern in the disc emission. No such structure is observed in
the quiescent disc (Marsh \& Horne 1990).

These observations provide a potential new constraint
on the angular momentum transport processes operating in accretion
discs. Two mechanisms are known that can provide a source of viscosity
in ionized, non-self-gravitating accretion discs in binary systems;
turbulence driven by the non-linear development of the Balbus-Hawley
instability (Balbus \& Hawley 1991; Tout \& Pringle 1992; 
Stone et al. 1996; Brandenburg et al. 1996); and spiral waves
or shocks driven by the gravitational perturbation of the secondary
(Sawada, Matsuda \& Hachisu 1986; Spruit 1987; Rozyczka \& Spruit 1989; 
Savonije, Papaloizou \& Lin 1994). It is obvious that the second
scenario leads to a spiral pattern of disc emission, but even if
tidally induced shocks are unimportant for the angular momentum 
transport budget in the steady-state they 
might still be observable in outburst, when the enhanced viscosity 
forces the disc to expand into a region where the strength of the
tidal forces is greater (Papaloizou \& Pringle 1977; Lin \& Pringle 
1976). We note that although it is generally believed 
that the spiral shock mechanism is inefficient in the relatively cool 
discs found in cataclysmic variables (Livio 1994; Savonije, Papaloizou \& 
Lin 1994), there are considerable theoretical uncertainties
in both mechanisms, and additional observational input is
highly desirable.

In this Letter, we compare simulations of accretion disc evolution
with the observations of IP Peg. Our goal is to test whether viscous
disc simulations, which are predicated on the existence of an internal
origin for the disc viscosity, are consistent with the strong spiral
structure observed in the data. We describe our calculations in Section 2,
and present in Section 3 model Doppler tomograms for comparison with 
the observations. Section 4 summarises our conclusions, and outlines
the theoretical expectations for spiral structure in other disc
systems.

\section{SIMULATION OF AN OUTBURST IN IP PEGASI}

The parameters of the dwarf nova IP Peg are given, for example, by
Warner (1995). The masses of the white dwarf and secondary are
$1.02 \ M_\odot$ and $0.5 \ M_\odot$ respectively, and the orbital
period is 3.8 hr. The outburst magnitude is $\sim 2$ magnitudes, 
and the rise to maximum lasts for 1-1.5 days. Thermal disc instability
models for such outbursts have been studied in great detail (e.g. 
Cannizzo 1993; Lasota 1998; and references therein), here we construct
a simplified three-dimensional simulation that nevertheless reproduces
the main dynamical effects of the outburst.

\subsection{Numerical method}

We calculate the evolution of the disc using the smooth particle
hydrodynamics (SPH) code described in detail by Murray (1996,1998; for 
general descriptions of SPH see Benz 1990; Monaghan 1992), with 
minor modifications to extend the code to three dimensions. As 
compared to other hydrodynamics methods, SPH codes have a relatively
large but well-characterized shear viscosity (Murray 1996), they 
are thus suitable choices for modelling discs {\em if} there is
a local source of viscous stresses. 

The quiescent disc is set up by commencing the simulation with 
an annulus of particles near the predicted tidal truncation 
radius (Papaloizou \& Pringle 1977), and allowing this to
relax while injecting gas at a steady rate through the inner
Lagrange point $L_1$. Particles are accreted onto the primary
when they stray inside an accretion radius $R_{\rm wd} = 0.02 a$,
where $a$ is the binary separation. As we are primarily interested 
in the dynamics of the outer disc we impose a fixed sound speed $c_s$,
and a constant SPH smoothing length $h = 0.02 a$. Under these
conditions the kinematic viscosity $\nu$ of the code is given by,
\begin{equation}
 \nu = {1 \over 10} \alpha_{\rm SPH} c_s h
 \label{SPH_viscosity}
\end{equation}
where $\alpha_{\rm SPH} = 1$ is the usual SPH linear viscosity 
co-efficient (Monaghan 1992). We take $\beta_{\rm SPH} = 0$, and
set $c_s$ such that the Mach number in the outer regions of the
quiescent disc is ${\cal{M}} \approx 30$. The resultant Shakura-Sunyaev
viscosity parameter $\alpha_{\rm SS}$ (Shakura \& Sunyaev 1973),
defined by,
\begin{equation}
 \nu = \alpha_{\rm SS} {c_s^2 \over \Omega}, 
 \label{SS_viscosity}
\end{equation}
with $\Omega$ the angular velocity, is then around 
$\alpha_{\rm SS} \simeq 0.15$ in the outer disc (note that it is
{\em not} constant with radius). Around 30 binary orbital
periods of evolution are required to reach an approximate
steady-state, at which point there are close to 
$2 \times 10^4$ particles in the disc. This is sufficient
for good resolution in the disc plane, though in the cool
quiescent state the simulation has only limited resolution
in the vertical direction.

In outburst, the sound speed and Shakura-Sunyaev $\alpha_{\rm SS}$
parameter increase. For IP Peg, whose outbursts are of rather modest
amplitude as compared to other dwarf novae, it is straightforward
to mimic this behaviour numerically. Starting with the steady
quiescent disc, we instantaneously increase $\alpha_{\rm SPH}$ by a 
factor of 4, and simultaneously raise $c_s$ by a factor of
$\sqrt{2}$. This leads to a rise in the bolometric luminosity 
of the disc by a factor of $\approx 6$, and is consistent with 
a $\sim 2$ magnitude increase in luminosity caused by a suddenly
enhanced mass accretion rate through a Shakura-Sunyaev disc.

\subsection{Disc structure in quiescence and in outburst}

\begin{figure*} 
 \psfig{figure=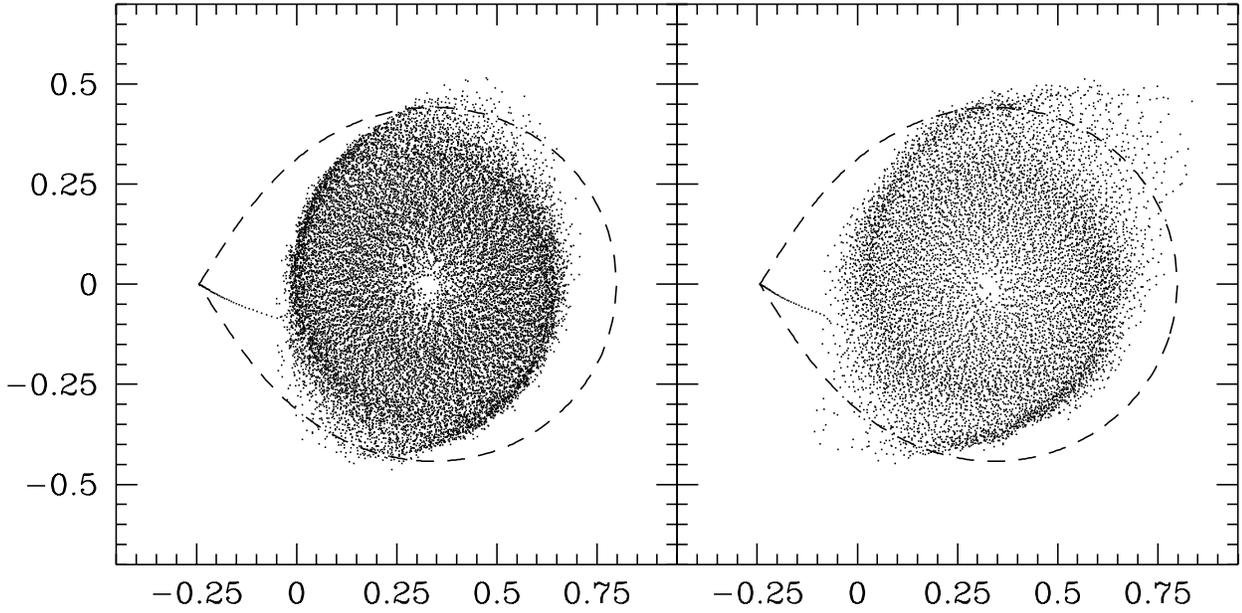,width=7.0truein,height=7.0truein}
 \vspace{-3.5truein}
 \caption{Simulations of the accretion disc of IP Peg in quiescence 
          (left panel) and outburst (right panel). The disc rotates
          counter-clockwise, with mass being added in an unresolved 
          stream from the $L_1$ point to the left of each panel. The
          dashed curve shows the Roche lobe of the primary. The
          disc in outburst is lower mass, somewhat more extended radially,
          and displays a clear two-armed trailing spiral structure
          in the outer regions.}
 \label{fig1}
\end{figure*}

Figure 1 shows the structure of the disc in quiescence and in outburst.
The disc in quiescence is mildly tidally distorted, with an outer radius
that varies from about $0.32 a$ to $0.4 a$. This is consistent with 
the expected size of a disc truncated at the outer edge by tidal torques
from the companion (Papaloizou \& Pringle 1997; Paczynski 1977). Clearly 
there is some non-axisymmetry in the outer part of the disc, but the
corresponding map of the disc luminosity (Figure 2) shows that the 
main non-axisymmetric contribution to the disc luminosity in quiescence 
arises from the hotspot or stream overflow region where the accretion 
stream from $L_1$ meets the disc. 
 
The right panels of Fig.~1 and Fig.~2 show the disc in the outburst
state. These snapshots were taken after the outburst had been in 
progress for 8 binary orbits, or around 30 hours for the parameters
of IP Peg. This approximately matches the timing of the 
observations of Steeghs et. al. (1997), which were taken a day
into outburst. By this stage in the simulation the disc mass had
declined by almost a factor of two -- comparable but probably
somewhat greater than would be expected in a detailed disc thermal
instability model.

In outburst, the disc is somewhat expanded relative to the quiescent
state and some material has escaped the Roche lobe of the primary -- 
large changes in the outer radius are not expected because
of the rapidly increasing strength of the tidal torques with increasing
radius. A two-armed spiral pattern in clearly seen in the outer 
regions of the disc, extending in phase from $\phi = 0.9-0.25$, 
and from $\phi = 0.4-0.75$. Fourier decomposition of the modes
present in the disc shows that this two-armed pattern is indeed
dominant, the resonant $m=1$ modes seen in simulations of low
mass ratio systems (Murray 1996) are relatively weak. The same
pattern is seen in the dissipation from the disc (Fig.~2), not 
only is the overall disc luminosity greatly increased in outburst,
but the integrated contribution from the spiral arms is now
comparable or greater than that from the hotspot in the
outer regions of the disc.
 
\begin{figure*} 
 \psfig{figure=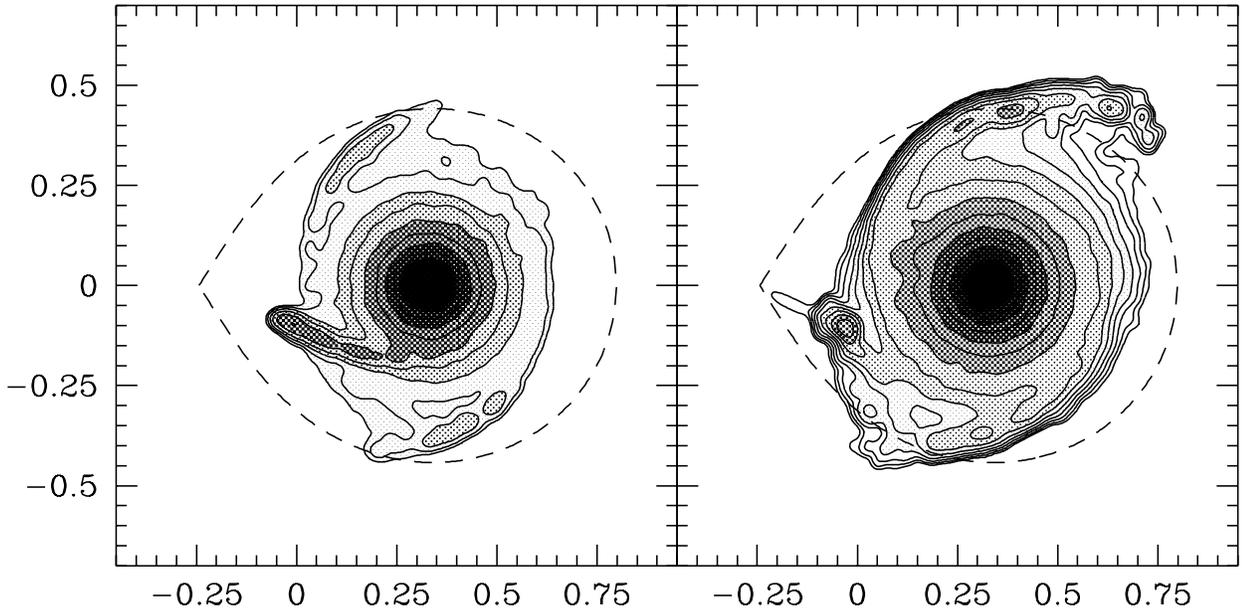,width=7.0truein,height=7.0truein}
 \vspace{-3.5truein}
 \caption{Luminosity of the disc in quiescence (left panel) and 
          outburst (right panel), at times corresponding to the 
          snapshots in Fig.~1. The contours and greyscale levels
          are the same in both plots, and are plotted at intervals
          of $\Delta \log (\sigma T_e^4) = 0.25$. The disc in
          outburst is hotter, and has a larger non-axisymmetric
          component to the emission in the outer regions.}
 \label{fig2}
\end{figure*}

\section{MODEL DOPPLER TOMOGRAMS}

\begin{figure*} 
 \psfig{figure=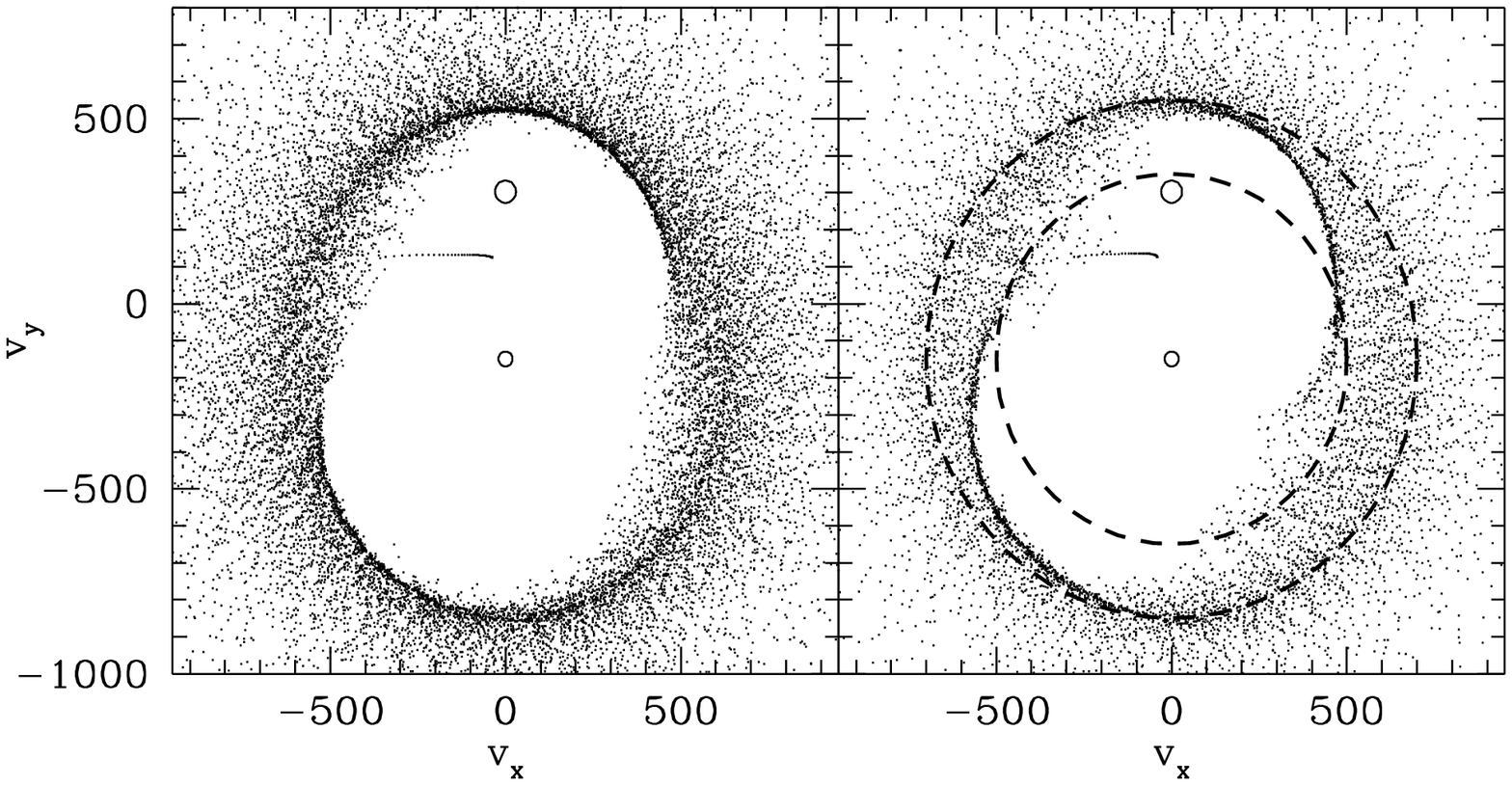,width=7.0truein,height=7.0truein}
 \vspace{-3.0truein}
 \caption{Particle positions from the simulation projected into 
          $(v_x, v_y)$ space in quiescence (left panel) and in
          outburst (right panel). The units are km/s. The small 
          circle shows the velocity of the white dwarf, the larger 
          circle that of the secondary. The dashed circles in the 
          outburst panel delineate the velocity extent of the spiral 
          structure, and are plotted at 500 km/s and 700 km/s
          relative to the velocity of the white dwarf.}
 \label{fig3}
\end{figure*}

The best observational probe of non-axisymmetry in interacting
binary accretion discs is provided by the technique of Doppler
tomography (Marsh \& Horne 1988). If the emission from the disc
is assumed stationary in the rotating frame of the binary, then 
the observed spectral line profiles as a function of phase $\phi$,
$f(v,\phi)$, can be inverted to yield the best-fitting 
{\em Doppler map}, $I(v_x,v_y)$, of line flux as a function 
of the velocity $\vec{v} = (v_x,v_y)$ in the binary frame. 
The detection of spiral structure in IP Peg by Steeghs et al.
(1997) is based on such a Doppler mapping method.

For a disc simulation, of course, both $\rho(v_x,v_y)$ and 
$I(v_x,v_y)$ (with some assumption about the disc emission) 
are directly available without the need for further assumptions.
In principle we can also construct the distorted maps that 
arise if the emission is not stationary in the binary frame,
although that is not required here as the two-armed spiral
pattern seen in the simulations evolved only slowly compared
to the orbital timescale. 

Figure 3 shows the disc density in quiescence and in outburst
projected into velocity space. The disc in quiescence is 
not axisymmetric in the outer regions, and this is reflected
in the distorted elliptical pattern seen at low velocities
in the quiescent disc. We note that during a lengthy period 
of quiescence the disc radius will decrease due to the addition 
of low angular momentum material from the accretion stream, and 
that this will reduce the strength of tidal distortions when 
compared to the (relatively short duration) simulations reported
here.

In outburst, the spiral arms seen in Fig.~1 are seen as clear
density enhancements in the upper right and lower left quadrants
of the map. These match the observed azimuthal extent of the
spiral arms seen in IP Peg. Also plotted are circles delineating 
the velocity range of the spiral features seen in the data, which 
were between approximately 500 km/s and 700 km/s relative to the 
white dwarf (Steeghs et al. 1997). Again this is in good agreement
with the velocity extent of the spiral arms seen in the simulation.

Figure 4 shows the simulated Doppler map for the quiescent and outburst states.
This is obtained by weighting the density map of Fig.~3 with the viscous
dissipation rate at the location of each particle, and then adaptively
smoothing the resulting map with a gaussian window function. Emission from the 
hot inner disc (which is noisy, and not seen in the H$\alpha$ and 
HeI line maps) has been removed by not plotting particles with 
velocities relative to the white dwarf of greater than 1000 km/s.
In general the outburst map matches the observations well, especially 
for the HeI line Doppler map. The same asymmetry between the 
strengths of the spiral arms is seen, and as the dominant emission
is found to arise from the high density regions in Fig.~3 the
velocity and azimuthal extent of the arms also agree well. The
hotspot emission is seen more clearly in the synthetic map 
than in the observations, suggesting that the luminosity 
from the hotspot is probably weak in the H$\alpha$ and HeI
lines.

\begin{figure*} 
 \psfig{figure=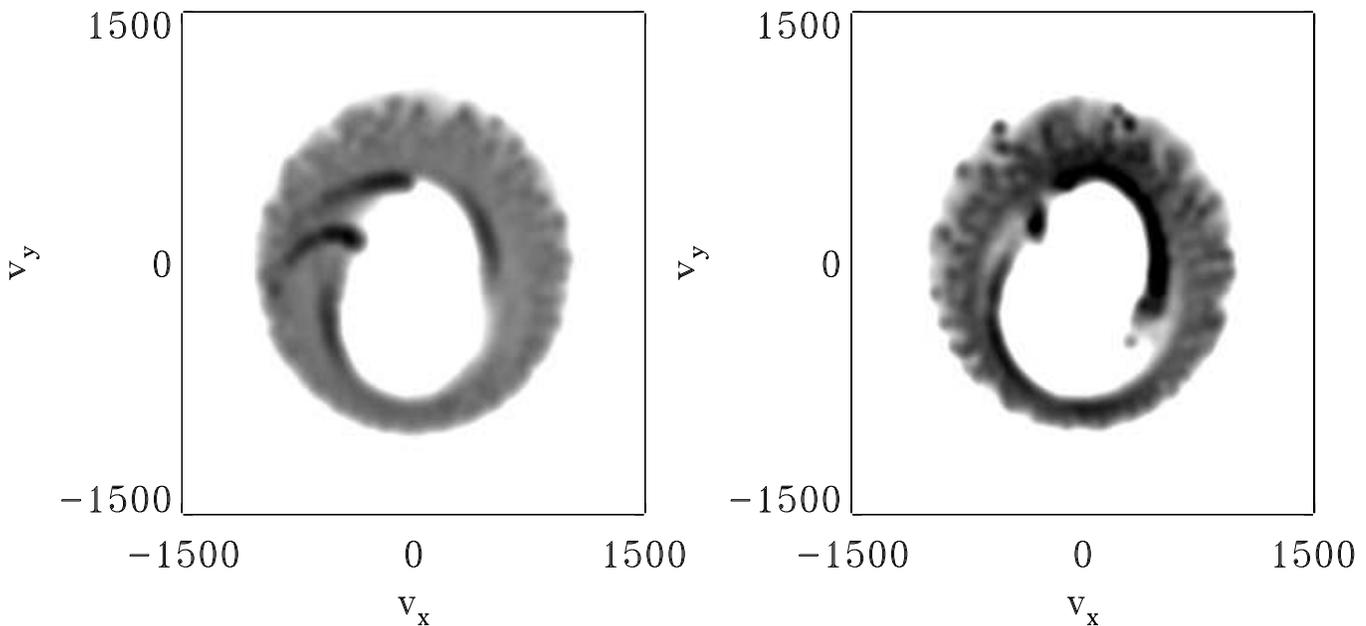,width=7.0truein,height=7.0truein}
 \vspace{-3.5truein}
 \caption{Map of dissipation projected into $(v_x, v_y)$ space
          for the disc during quiescence (left panel) and in
          outburst (right panel). The units are km/s. In each case
          the greyscale is linear, but the lowest intensity in 
          the quiescent map is a factor of 6 reduced from that in
          outburst. Note that in outburst the
          spiral arm in the upper right quadrant is markedly
          stronger than that in the lower left quadrant of the map.}
 \label{fig4}
\end{figure*}

As a check against possible systematic errors arising from 
the Doppler mapping procedure, we have also generated maps
indirectly by first creating synthetic trailed spectra from
the simulation, and then projecting the spectra into the 
Doppler map using the same techniques as are used observationally.
The map seen in Fig.~4 is again recovered, though at the
cost of considerably enhanced noise.

\section{DISCUSSION}

In this Letter we have presented a simulation of the evolution of
the accretion disc for the parameters of the binary IP Pegasi
in outburst. We find as our main result that a
spiral pattern is formed in the outer disc during outburst as the
enhanced viscous stresses push the disc edge into a region of strong
gravitational perturbations from the secondary. The spiral structure
obtained in the simulation is two-armed, non-resonant, and much
more prominent in outburst as compared to quiescence. Comparing 
the results with the observations of Steeghs, Harlaftis \& Horne
(1997), we find that there is excellent agreement with the azimuthal 
extent, velocity range, and asymmetry of the observed pattern.

The observations of IP Pegasi and other cataclysmic variables find
no clear evidence for spiral patterns in quiescent discs. This is 
consistent with theoretical expectations provided that the quiescent 
discs are cool, viscous, and have not expanded close to the tidal 
radius. Observations following the decay of spiral structure 
after the outburst has ended would be valuable in understanding the interplay 
of these factors. However the current evidence continues to support
the conclusion of Savonije, Papaloizou \& Lin (1994) 
that angular momentum transport by spiral shocks is insufficiently
effective to provide the bulk of the angular momentum transport
in the relatively cool, thin discs found in cataclysmic variables. 
Spiral shocks {\em are} likely to be important over a wide range
of radii in the hotter discs found in X-ray binaries (Owen \&
Blondin 1997), and in the quasi-spherical accretion flows postulated
for the advectively dominated regime, although internal sources 
of viscosity are also likely to be more efficient in those thicker
disc geometries.

The current observations can be modelled adequately using a highly
simplified three-dimensional model of an outburst caused by a thermal 
disc instability, in which the only inputs are the change in $\alpha_{\rm SS}$
and $c_s$ between quiescence and outburst. Future observations,
extending over the rise to outburst and during the decline, may 
be able to provide stronger constraints on the assumed disc model.
In particular, since the spiral pattern arises as a result of 
the imbalance between internal viscous stresses and well-understood
gravitational torques, such observations can probe the variation 
of the disc viscosity with the local physical conditions in the
disc. Eclipsing systems such as IP Peg are particularly promising in
this regard as the radial run of quantities such as the effective 
temperature can readily be derived simultaneously from eclipse
mapping.

The detection of spiral structure in the accretion disc of IP Peg,
which has relatively feeble outbursts, implies that similar or stronger
features may be expected in most dwarf novae. Theoretically, we
note that qualitative differences are expected in systems with 
low mass ratio (roughly, $q < 1/4$). In these binaries both the observations 
of superhumps in the light curve, and numerical simulations (Murray 
1996, 1998), suggest that a strong $m=1$ mode is excited in outburst.
These features are {\em not} stationary in the corotating binary frame,
making detailed investigation via Doppler mapping harder.
It would also be worthwhile investigating whether spiral structure
is observable in magnetic systems such as intermediate polars, where
non-axisymmetry might be induced at the {\em inner} edge of the 
disc as a result of magnetic torques from the white dwarf.

{\em Note added:} Godon, Livio \& Lubow (1998) have recently 
presented calculations showing that a steady tidally induced spiral pattern 
does not match the observations of IP Peg. This is consistent with our
finding that consideration of the transient behaviour of the disc during 
outburst is required.

\end{document}